\documentclass[aps,prl,superscriptaddress, reprint]{revtex4-2}
\usepackage{CJK}
\usepackage{braket}
\usepackage{amsmath}
\usepackage{amssymb}
\usepackage{graphicx}
\usepackage{float}
\usepackage{hyperref}
\usepackage{xcolor}
\usepackage{verbatim}
\usepackage[normalem]{ulem}
\newcommand{\R}{\tilde{\rho}}
\newcommand{\Rss}{\tilde{\rho}^{ss}}

\newcommand{\sv}[1]{}

\begin{document}
\begin{CJK*}{GB}{}
\newcommand{\floor}[1]{\lfloor #1 \rfloor}
\title{Cooperation and Competition in Synchronous Open Quantum Systems}
\author{Taufiq Murtadho}
\affiliation{Center for Theoretical Physics of Complex Systems, Institute for Basic Science (IBS), Daejeon 34126, Republic of Korea.}
\affiliation{Basic Science Program, Korea University of Science and Technology, Daejeon 34113, Republic of Korea.}
\affiliation{School of Physical and Mathematical Science, Nanyang Technological University, Singapore 639798, Singapore}
	\author{Sai Vinjanampathy}
\email[]{sai@phy.iitb.ac.in}
\affiliation{Department of Physics, Indian Institute of Technology-Bombay, Powai, Mumbai 400076, India.}
\affiliation{Centre of Excellence in Quantum Information, Computation, Science and Technology, Indian Institute of Technology Bombay, Powai, Mumbai 400076, India.}
\affiliation{Centre for Quantum Technologies, National University of Singapore, 3 Science Drive 2, Singapore 117543, Singapore.}

\author{Juzar Thingna}
\email[]{juzar_thingna@uml.edu}
\affiliation{Center for Theoretical Physics of Complex Systems, Institute for Basic Science (IBS), Daejeon 34126, Republic of Korea.}
\affiliation{Basic Science Program, Korea University of Science and Technology, Daejeon 34113, Republic of Korea.}
\affiliation{Department of Physics and Applied Physics, University of Massachusetts, Lowell, MA 01854, USA.}
\date{\today}
\begin{abstract}
    Synchronization between limit cycle oscillators can arise through entrainment to an external drive or through mutual coupling. The interplay between the two mechanisms has been studied in classical synchronizing systems, but not in quantum systems. Here, we point out that competition and cooperation between the two mechanisms can occur due to phase pulling and phase repulsion in quantum systems. We study their interplay in collectively driven degenerate quantum thermal machines and show that these mechanisms either cooperate or compete depending on the working mode of the machine (refrigerator or engine). The entrainment-mutual synchronization interplay persists with an increase in the number of degenerate levels, while in the thermodynamic limit of degeneracy, mutual synchronization dominates. Overall, our work investigates the effect of degeneracy and multilevel scaling of quantum synchronization and shows how different synchronizing mechanisms can cooperate and compete in quantum systems.
\end{abstract}
\maketitle
\end{CJK*}
\noindent \textit{Introduction.--} Synchronization is a ubiquitous phenomenon in which stable phase relations emerge between multiple limit cycle oscillators \cite{pikovsky2002synchronization}. There are two main mechanisms that give rise to synchronization: i. \textit{Entrainment} that refers to synchronization of an oscillator by unidirectional coupling to a periodic external drive \cite{adler1946astudy}, and ii. \textit{Mutual synchronization} which refers to the adjustment of rhythms of two or more mutually coupled oscillators, such as in the widely-known Kuramoto model \cite{acebron2005the}. These two mechanisms may coexist in some systems  \cite{sakaguchi1988cooperative, childs2008stability, snyder2017stability, goldobin2017competing,sakaguchi1988cooperative, childs2008stability} and their interplay has also been experimentally studied in globally coupled electrochemical oscillators \cite{kiss2008resonance}.
 
In the same spirit as classical synchronization, quantum synchronization is often studied through entrainment \cite{walter2014quantum,lee2013quantum, chia2020relaxation, roulet2018synchronizing,  ben2021quantum, koppenhofer2019optimal} or mutual coupling \cite{lee2014entanglement, xu2014synchronization, walter2015quantum, zhu2015synchronization,weiner2017phase, roulet2018quantum, FinnPRL22} and has been experimentally observed recently \cite{koppenhofer2020quantum,laskar2020observation,krithika2022observation}. However, unlike classical synchronization, the coexistence and the interplay between these two mechanisms in the quantum regime has not been investigated. Understanding this interplay is crucial in the control of various quantum technologies where both driving and interaction are important such as in superradiant lasers \cite{xu2014synchronization}, coupled time-crystals \cite{hajdusek2022seeding}, and coupled heat engines \cite{herpich2018collective, jaseem2020quantum, ryu2021stochastic, solanki2022role}. 

In this work, we show that the phases of steady-state coherences follow a phase synchronization model, where the external entraining drive competes with the mutually coupled phases. This opens up the possibility of observing well-studied classical phenomena, such as synchronization-anti-synchronization transition~\cite{KIM200339} and chimera \cite{KuramotoNon02, AbramsPRL04}, in the quantum regime. Our framework applies to generic quantum systems, with one or more external drives that couple the coherences that themselves are mutually coupled, either coherently or dissipatively, leading to an interplay between entrainment and mutual synchronization.  

As a concrete example, we consider a degenerate multilevel generalization of the Scovil--Schulz-DuBois maser heat engine~\cite{scovil1959three}, where the external collective drive connects transitions between the degenerate manifold and the first-excited state~\cite{dorfman2018efficiency}. The states within the degenerate manifold mutually interact to form a \textit{stable } collective symmetric (in-phase) and anti-symmetric (out-of-phase) superposition (mutual synchronization). At the same time, the external drive causes the phases within the degenerate manifold to be aligned in-phase with the drive (entrainment). In the engine regime, stimulated emission consumes the collective symmetric superposition state thereby enhancing the population of the \emph{anti-symmetric} state. Thus, there is competition between entrainment (in-phase) and mutual synchronization (out-of-phase). In the refrigerator regime, the stimulated absorption enhances the population of the collective \emph{symmetric} superposition state thereby always cooperating with entrainment. Our work sheds light on the synergistic interplay between entrainment and mutual synchronization in quantum systems.

\noindent \textit{Quantum synchronization in $D$-level systems.--} Quantum synchronization has been studied in systems with continuous degrees of freedom such as oscillators~\cite{lee2013quantum,MariPRL13,walter2014quantum,lee2014entanglement,walter2015quantum,chia2020relaxation,ben2021quantum} and discrete degrees of freedom such as spin-1 systems \cite{roulet2018synchronizing, roulet2018quantum, koppenhofer2019optimal}. A wide variety of measures, based on various physical and mathematical motivations such as phase-space based measures \cite{walter2014quantum, roulet2018quantum, roulet2018synchronizing}, correlation measures \cite{Galve2017}, and information-theoretic measures~\cite{{jaseem2020generalized}} has been used to quantify synchronization.

In this work, we use the phase-space based measure built on the Husimi-Q phase space representation \cite{schleich2011quantum, tilma2011sun} of the steady-state $\rho^{ss}$ with respect to $SU(D)$ coherent state~\cite{nemoto2000generalized, tilma2011sun} defined as 
\begin{equation}
    \label{eq:husimiQ}
    Q[\rho^{ss}] = \frac{D!}{\pi^{D-1}}\braket{\alpha_D|\rho^{ss}|\alpha_D},
\end{equation}
where  $\ket{\alpha_D} = \sum_{n = 1}^{D}\alpha_{n}\ket{n}$ is the $SU(D)$ coherent state with coefficients 
\begin{equation}
    \label{eq:SU(D)}
    \alpha_n = \begin{cases}
    e^{i\phi_{n}}\cos\theta_{n}\prod_{k =1}^{n-1} \sin\theta_{k} & 1 \leq n< D\\
    e^{i\phi_{D}}\prod_{k=1}^{D-1}\sin\theta_k & n = D,
    \end{cases} 
\end{equation}
where it is implicitly assumed that the product term is identity for $n=1$ and the reference phase $\phi_1 = 0$. The synchronization measure is given by the difference between integrating out the angles $\theta_k$ corresponding to the population degrees of freedom and doing the same for the uniform measure, given by
\begin{align}
    \label{eq:quasidistribution}
    S(\phi_1, \cdots, \phi_{D-1}) &= \int Q[\rho^{ss}] d\Theta -\frac{1}{(2\pi)^{D-1}}\nonumber\\
   & = \frac{1}{2^{D+1}\pi^{D-2}}\sum_{n\neq m}\rho_{nm}^{ss}e^{i(\phi_m-\phi_n)},
\end{align}
which lives on a $D-1$ dimensional torus (see~\hyperref[appendixA]{Append. A}). The distribution $S(\phi_1,\cdots,\phi_{D-1})$ is zero everywhere for a diagonal steady-state which is interpreted as a limit cycle~\cite{jaseem2020generalized} possessing stable amplitudes (fix diagonal elements) but free phases. The notion of free-phase in a such diagonal limit cycle is analogous to a classical stochastic limit cycle whose phase distribution approaches a uniform distribution in the steady-state~\cite{pikovsky2002synchronization, kurrer1991effect, koppenhofer2019optimal, ben2021quantum, eneriz2019degree}. 

We associate the peak of $S(\phi_1,\cdots, \phi_{D-1})$ as a phase-space synchronization measure~\cite{roulet2018synchronizing, roulet2018quantum,eneriz2019degree},
\begin{equation}
    \label{eq:smax_formula}
    S_{max} =  \max_{\phi_1, \cdots, \phi_{D-1}}\frac{1}{2^{D+1}\pi^{D-2}}\sum_{n\neq m}\rho_{nm}^{ss}e^{i(\phi_m-\phi_n)}.
\end{equation}
The synchronization measure, $S_{max}$ only depends on the steady-state coherences. However, we note that a high value of $S_{max}$ requires all phase preferences $\Phi_{ij} = \arg(\rho_{ij}^{ss})$ to be compatible, i.e., $\Phi_{ij} - \Phi_{jk} = \Phi_{ik}$ $\forall i
\neq  j \neq k$, a condition that is stronger than the mere presence of coherences.

\noindent \textit{Degenerate thermal maser.--}
\begin{figure}
    \centering
    \includegraphics[width = 0.8\columnwidth]{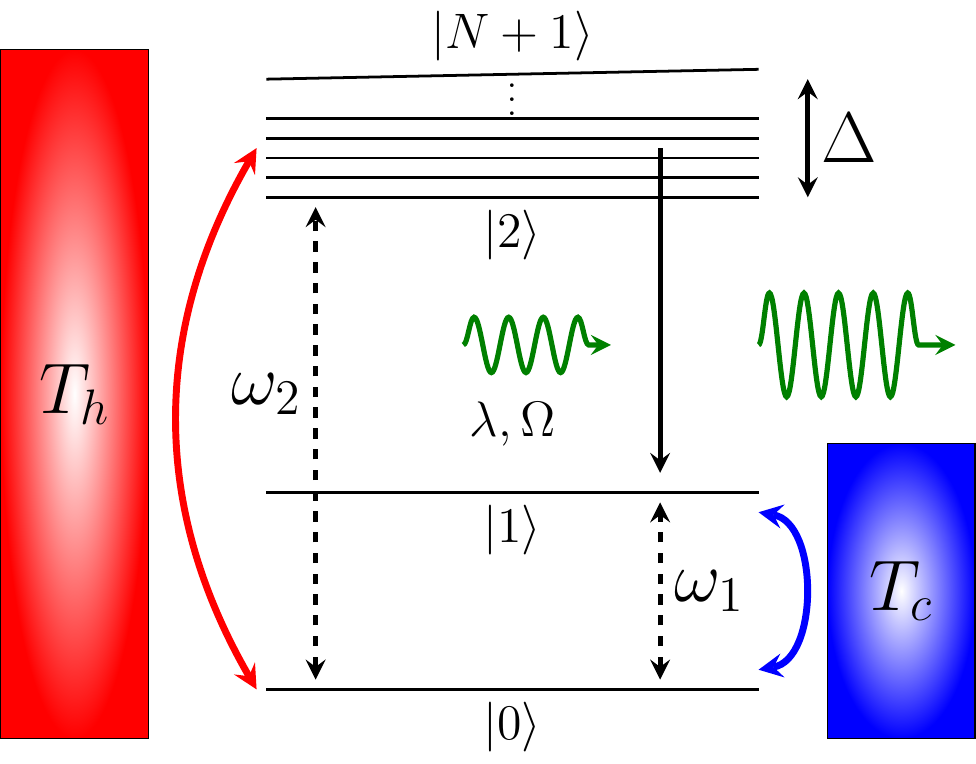}
    \label{fig1}
    \caption{Schematic of the degenerate quantum thermal maser, which is a generalization of the standard Scovil--Schulz-DuBois three-level thermal maser \cite{scovil1959three}. Here, $N$ is the number of states in the degenerate manifold and here we focus on the case $\Delta = 0$. The near-degenerate case where $\Delta \neq 0$ is discussed in the accompanying manuscript [\onlinecite{MurtadhoPRA23}]}
\end{figure} 
Entrainment in quantum systems is the result of an interplay between coherent driving and dissipation \cite{roulet2018synchronizing, lee2013quantum}. The system we consider is depicted in Fig.~\ref{fig1} and consists of $(N+2)$ levels whose bare Hamiltonian is given by,
\begin{equation}
    \label{eq: bare_Hamiltonian}
    H_0 = \omega_1 \ket{1}\bra{1}+ \sum_{j=2}^{N+1}\omega_j\ket{j}\bra{j},
\end{equation}
with $\omega_{j+1}>\omega_j,\omega_0=0$. The upper $N$ levels are degenerate with $\omega_2 = \omega_3 = \cdots = \omega_{N+1}$. Although we work in the limit of exact degeneracy, our main results hold even in the near-degenerate scenario and will be considered in detail in an accompanying Ref. [\onlinecite{MurtadhoPRA23}]. 

This system is driven by a monochromatic drive $V(t)$ of frequency $\Omega$ given by
\begin{align}
    \label{eq: drive_form}
    V(t) = \sum_{j=2}^{N+1}\lambda_j e^{i\Omega t} \ket{1}\bra{j} + \text{h.c.}
\end{align}
This drive can be rewritten as a coupling to a collective-transition mode $\ket{1} \leftrightarrow \ket{J} = (1/\lambda_{eff})\sum_{j}\lambda_j\ket{j}$ with $\lambda_{eff} = \sqrt{\sum_{j}|\lambda_j|^2}$ being the effective coupling strength. Such collective drives are realizable in an ensemble of atoms driven by light, if the inter-atomic distance is much smaller than the wavelength of the light, such as in the case of Dicke superradiance \cite{gross1982superradiance}.

The system is acted upon by a dissipator
\begin{equation}
    \label{eq:dissipator}
    \mathcal{D}[\rho] =  \sum_{\mu =1}^{2}\Big[\Gamma_{c_\mu}\mathcal{L}[c_{\mu}]\rho+\sum_{j=2}^{N+1}\Gamma_{h_\mu}\mathcal{L}[h_{\mu}^{j}]\rho\Big],
\end{equation}
which leads to a multilevel generalization of the Scovil--Schulz-DuBois maser heat engine~\cite{scovil1959three,dorfman2018efficiency}. The dissipator $\mathcal{L}[X]\rho = 2X\rho X^\dagger - \{X^\dagger X, \rho\}$ is of the Lindblad form such that the hot (cold) bath with jump operators $h_1^{j} = h_2^{j\dagger} = \ket{0}\bra{j}$ ($c_1 = c_2^\dagger = \ket{0}\bra{1}$) induce transitions between the ground state and the degenerated manifold (first-excited state). The associated rates follow local-detailed balance and are given by $\Gamma_{h_{1}(c_1)} = \gamma_{h(c)}(1+n_{h(c)})$ and $\Gamma_{h_2(c_2)} = \gamma_{h(c)}n_{h(c)}$ with $\gamma_{h(c)}$ being the effective system-bath coupling strength and $n_{h(c)} = \left[\exp(\beta_{h(c)}\omega_{2(1)})-1\right]^{-1}$ being the Bose-Einstein distribution at inverse temperature $\beta_{h(c)}$. The action of the heat baths leads to a population inverted steady state between the first-excited state $|1\rangle$ and the degenerated manifold \{$|j\rangle$, $\forall j=2,\cdots, N+1$\} if $n_h>n_c$. If there is population inversion, the system behaves as a maser heat engine \cite{boukobza2007three}. However, if $n_h<n_c$, population inversion is lost and the system behaves as a refrigerator by attenuating the drive \cite{boukobza2007three}. We can rewrite the Hamiltonian in a frame co-rotating with the drive as $\tilde{H} = ({\Omega}/{2})(\sum_{j=2}^{N+1}\ket{j}\bra{j} - \ket{1}\bra{1})$ giving us the rotating frame quantum master equation,
\begin{equation}
\label{eq:rot_QME}
\frac{d\tilde{\rho}}{dt} = -i[H_0-\tilde{H}+\tilde{V},\tilde{\rho}]+\mathcal{D}[\tilde{\rho}],
\end{equation}where $\tilde{O} \equiv  e^{-i\tilde{H} t}O e^{i\tilde{H} t}$ ($O=\rho,V$) is an operator in the rotated frame with $\tilde{V} = \sum_{j=2}^{N+1}\lambda_j\ket{1}\bra{j} + \mathrm{h.c.}$.

\begin{figure}
    \centering
    \includegraphics[width=0.9\columnwidth]{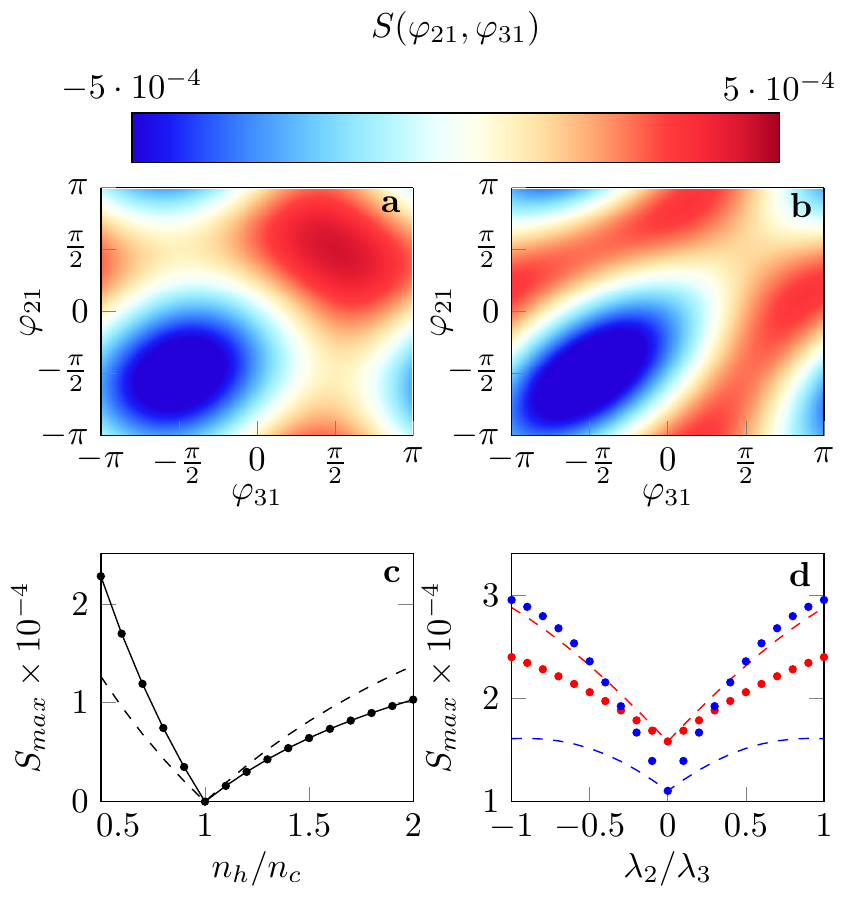}
    \caption{Interplay between entrainment and mutual coupling for $N=2$. Panels {\bf a} and {\bf b} show phase quasi-distribution function $S(\varphi_{21}, \varphi_{31})$ [Eq.~\eqref{eq:quasidistribution}] where $\varphi_{ij} = \phi_i - \phi_j$ in the engine regime ($n_h/n_c$ = 100). For $k = 3$, $S(\varphi_{21}, \varphi_{31})$ shows a localized maximum when the phases are in-phase ($\varphi_{21}-\varphi_{31} \approx 0$ in the red-region in {\bf a}, entrainment-dominant). Whereas for $k = 0.75$, when $S(\varphi_{21}, \varphi_{31})$ is maximized the phases do not localize but their difference is out-of-phase ($\varphi_{21}-\varphi_{31} \approx \pi$ in the red-region in {\bf b}, mutual coupling dominant). Panel {\bf c} shows $S_{max}$ (solid circle) as a function of $n_h/n_c$ with the solid line representing the analytic prediction of Eq.~\eqref{eq:smax_result}. The dashed line is the entrainment contribution to $S_{max}$, i.e., $(|\rho_{12}|+|\rho_{13}|)/16\pi^2$. The vertical dotted line represents the boundary between refrigerator ($n_h/n_c<1$) and engine ($n_h/n_c>1$) regimes. Panel {\bf d} shows $S_{max}$ (solid circle) and $(|\rho_{12}|+|\rho_{13}|)/16\pi^2$ (dashed line) plotted against inhomogeneous driving strength ratio $|\lambda_2/\lambda_3|\leq 1$ in the engine (red) and refrigerator (blue) regimes indicating competition (cooperation) between entrainment and mutual coupling is robust in the engine (refrigerator) regime. The other parameter values are $\omega_2 = \omega_3 = 3\omega_1,\; \Omega = \omega_2 - \omega_1, \; \gamma_c = 0.2\omega_1, \; \gamma_h = 0.05\omega_1, \; n_c = 0.5$, and $\lambda_2 = 0.1\omega_1$}
     \label{fig2}
\end{figure}

\noindent \textit{Competition vs cooperation.--} Equation~\eqref{eq:rot_QME} can be solved analytically for the case of homogeneous driving strength $\lambda_j = \lambda$ ($\forall j=2,\cdots, N+1$) and resonant driving $\Omega = \omega_2 - \omega_1$. In this case, the steady-state coherences are given by
\begin{eqnarray}
    \label{eq:ss_nondeg_coherences}
\tilde{\rho}_{1j}^{ss} &=& i\frac{\lambda(n_c-n_h)\gamma_c\gamma_h(1+n_h)}{F(N,n_h, n_c, \gamma_c, \gamma_h, \lambda)},\\
    \label{eq:ss_deg_coherences}
		\tilde{\rho}_{jl}^{ss} &=& \frac{\lambda^2\gamma_c(n_c - n_h)}{F(N,n_h, n_c, \gamma_h, \gamma_c, \lambda)},
\end{eqnarray}
where $j,l = 2,\cdots,N+1$, $j\neq l$ and the function $F(N,n_h, n_c, \gamma_c, \gamma_h, \lambda) = AN^2+BN+C$ with $A,B,$ and $C$ being positive constants that depend on all remaining parameters (see \hyperref[appendixB]{Append. B} for the explicit expressions for these constants).

The \textit{non-degenerate coherences} ($\tilde{\rho}_{1j}$) are directly induced (i.e., $\propto \lambda$) by the drive whereas the \textit{degenerate coherences} ($\tilde{\rho}_{jl}$) are an indirect consequence ($\propto \lambda^2$) of the collective nature of the drive. Their differences are clear as one transforms back to the original frame in which $\rho_{1j}= \tilde{\rho}_{1j}e^{-i\Omega t}$ and $\rho_{jl} = \tilde{\rho}_{jl}$. The phase preferences induced by $\rho_{1j}$ rotate with the driving frequency while that of $\rho_{jl}$ remain stationary in the original frame. Both of these coherences affect the phase distributions of the states within the degenerate manifold. For these reasons, we infer that there are two synchronization mechanisms at play in this system, entrainment induced directly by the drive and mutual coupling that occurs due to the presence of a degenerate manifold. Entrainment induces phases relative to driving whose effect is the emergence of stable non-degenerate coherences $\tilde{\rho}_{1j}^{ss}$. On the other hand, mutual coupling induces a relative phase between states in the degenerated manifold independent of the driving phase, which is reflected by stable degenerate coherences $\tilde{\rho}_{jl}^{ss}$.

Recall that we have denoted $\Phi_{ij} = \arg(\tilde{\rho}_{ij}^{ss})$ as the steady-state phase preferences. When there are multiple of such preferences, synchronization requires all the phase relations to be compatible, i.e. $\Phi_{ij} - \Phi_{jk} = \Phi_{ik}$ ($i \neq j \neq k$). However, we find that in our system such a condition is only satisfied in the refrigerator regime where $\Phi_{1j} = \pi/2$~($\forall j$) and $\Phi_{jl} = 0$~($j\neq l$). In the engine regime, we have $\Phi_{1j} = -\pi/2$ ($\forall j$) and yet $\Phi_{jl} = \pi$~($j\neq l$). We interpret this as a result of an interplay between entrainment and mutual coupling. We find that entrainment always pulls the degenerate states to be \textit{in-phase} (Fig. \hyperref[fig2]{2{\bf a}}). Mutual coupling prefers \textit{out-of-phase} configuration in the engine regime (Fig. \hyperref[fig2]{2{\bf b}}), and in-phase configuration in the refrigerator regime. Consequently, we expect entrainment and mutual coupling to cooperate in the refrigerator regime and compete in the engine regime.

\begin{figure}
    \centering
    \includegraphics[width = 0.9\columnwidth]{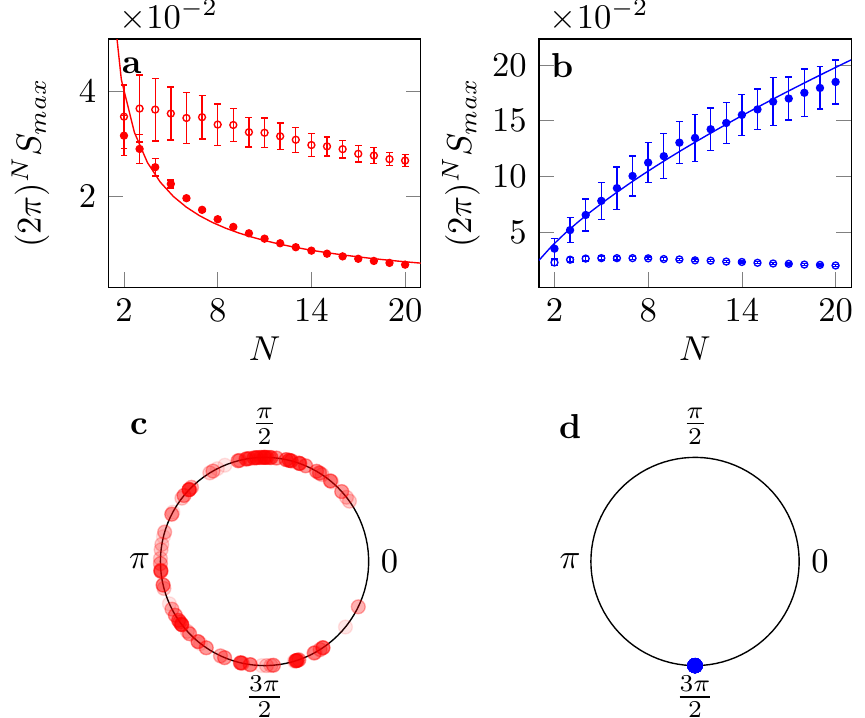}
    \caption{Panels \textbf{a-b} show $\mathbb{S}_{max} = (2\pi)^N S_{max}$ (solid circle) compared with its entrainment contribution $\frac{1}{4}\sum_{j=2}^{N+1}|\rho_{1j}|$ (empty circle) as a function of $N$. The error bar is calculated from $10^2$ random realizations of driving strength ratio $\lambda_j/\lambda_2\leq 1$ for $\lambda_j\geq 0$ and $j =3,\cdots,N+1$ with $\lambda_2$ held constant. The solid lines are the curve fits using $\mathbb{S}_{max} \propto N^{\alpha}$ with $\alpha = -0.72$ ({\bf a}) and $=0.69$ ({\bf b}). Panels \textbf{c-d} show optimum phases $\{\varphi_{j1}^{opt}\}$ for $N=20$ in in the engine (\textbf{c}, $n_h/n_c = 10$) and refrigerator (\textbf{d}, $n_h/n_c = 0.4$) regimes plotted on a unit circle. The different opacity represents different realizations of $\lambda_j/\lambda_2$ ($j =3,\cdots,N+1$). All the phases in all realizations coalesce to a single data point in the refrigerator case. All other parameters are the same as Fig.~\ref{fig2}.} 
    \label{fig3} 
\end{figure}

The competition and cooperation are obvious when we calculate the phase space synchronization measure $S_{max}$ [see Eq. \eqref{eq:smax_formula}]. In general, this requires optimization over $N$ variables which we calculate analytically for $N=2$ (see \hyperref[appendixC]{Append. C})
\begin{equation}
	\label{eq:smax_result}
		S_{max}=\dfrac{1}{16\pi^2}\times \begin{cases}
		|\tilde{\rho}_{12}^{ss}|+|\tilde{\rho}_{13}^{ss}|+|\tilde{\rho}_{23}^{ss}|&{\scriptstyle \mathrm{if} \;  n_h < n_c}\\
        |\tilde{\rho}_{12}^{ss}|+|\tilde{\rho}_{13}^{ss}|-|\tilde{\rho}_{23}^{ss}| & {\scriptstyle \mathrm{if}\; n_h > n_c \; \& \; k>2}\\
        \left(1+\dfrac{k^2}{2}\right)|\tilde{\rho}_{23}^{ss}| & {\scriptstyle \mathrm{if}\; n_h>n_c \; \&\;  k<2},
		\end{cases}
\end{equation}
where $k = \gamma_h(1+n_h)/\lambda = |\tilde{\rho}_{12}^{ss}|/|\tilde{\rho}_{23}^{ss}| = |\tilde{\rho}_{13}^{ss}|/|\tilde{\rho}_{23}^{ss}|$ is the \textit{dissipation-to-driving} ratio. The set of optimal phases $(\varphi_{21}^{opt}, \varphi_{31}^{opt})\equiv (\varphi_{21}, \varphi_{31})|_{S = S_{max}}$ evaluated in~\hyperref[appendixC]{Append. C} are given by,
\begin{equation}
    \label{eq:optimum_phases}
    (\varphi_{21}^{opt}, \varphi_{31}^{opt})= \begin{cases}
    	\left(-\dfrac{\pi}{2},-\dfrac{\pi}{2}\right)&{\scriptstyle \mathrm{if} \; n_h<n_c}\\
        \left(\dfrac{\pi}{2}, \dfrac{\pi}{2}\right)&{\scriptstyle \mathrm{if}\; n_h > n_c \; \& \; k>2}\\
        (\chi, \pi-\chi)\; \&\; (\pi-\chi, \chi)  &{\scriptstyle \mathrm{if}\; n_h>n_c \; \&\;  k<2},
    \end{cases}
\end{equation}
where $\varphi_{ij} = \phi_{i} - \phi_{j}$ and $\chi = \arcsin(k/2)$. Equations \eqref{eq:smax_result}-\eqref{eq:optimum_phases} show the effect of the coherent drive and bath couplings on the synchronous dynamics of the system. Cooperation in the refrigerator regime ($n_c > n_h$) is reflected by the fact that each component of the magnitude of coherence adds up in the synchronization measure $S_{max}$, whereas in the engine case there is competition since the mutual coupling component $|\rho^{ss}_{23}|$ reduces the effect of the entrainment contribution $|\rho^{ss}_{12}|+|\rho^{ss}_{13}|$. %The phases coupled incoherently via interactions with a common bath will tend to align themselves, whereas the situation is subtle when the external drive competes with the coupled phases, resulting in a solution that balances entrainment and mutual coupling. 
In other words, the phases are either equal in some cases or they are arranged antipodally in other cases, as shown in Eq.\eqref{eq:optimum_phases}.

In the engine regime, $S_{max}$ is also divided into regimes where entrainment is dominant ($k>2$) and where the mutual coupling is dominant ($k<2$). For the entrainment dominant regime, the competition is apparent from the negative contribution of $|\rho_{23}^{ss}|$ to $S_{max}$. Note that this is different from the previously reported phenomenon of synchronization blockade \cite{koppenhofer2019optimal,solanki2022symmetries}, in our case, $S_{max}$ can not vanish except for $\lambda = 0$ or $n_h = n_c$ where the steady-state is diagonal (see \hyperref[appendixD]{Append. D}). The transition from entrainment to mutual coupling dominant regime is shown in Figs.~\hyperref[fig2]{2{\bf a}-{\bf b}} where we plot the phase distribution $S(\varphi_{21}, \varphi_{31})$ for different $k$ values. In particular, we see that as we cross $k=2$, the relative phases go from in-phase to out-of-phase. Moreover, the localization pattern changes from a point localization to ring localization (on a torus), wherein the latter only the relative phase $\varphi_{23} = \varphi_{21} - \varphi_{31}$ is fixed, indicating that entrainment is lost.

The competition and cooperation observed is also robust with respect to all values of individual driving strength ratio $\lambda_2/\lambda_3$ as shown in Fig.~\hyperref[fig2]{2{\bf d}}. Interestingly, $S_{max}$ is symmetric with respect to a transformation $\lambda_j \rightarrow -\lambda_j$ which transforms $\tilde{\rho}_{jl}^{ss}\rightarrow -\tilde{\rho}_{jl}^{ss}$ for all $l\neq j$. This can be intuitively explained by $S_{max}$ only depending on the norm of coherences. In this case, the phase preference of entrainment and mutual coupling is reversed, i.e. both prefer \textit{out-of-phase} in the refrigerator regime while mutual coupling (entrainment) prefers \textit{in-phase} (\textit{out-of-phase}).  

\noindent \textit{Scaling with $N$.--} Calculating $S_{max}$ boils down to performing $N$-variable optimization which in general is difficult for $N>2$. However, in the refrigerator regime, assuming homogeneous driving $\lambda_j = \lambda$ the problem simplifies and one can show that $S(\{\varphi_{1j}\})$ saturates the $l_1$-norm bound~\cite{jaseem2020generalized} (see \hyperref[appendixA]{Append. A} for a proof). Thus, we conclude that in the refrigerator regime $S_{max} \propto C_{l_1} = \sum_{i< j}^{N+1}|\tilde{\rho}_{ij}|$ for any $N$. 

The fact that $S_{max}$ is always proportional to the $l_1$-norm in the refrigerator regime demonstrates that entrainment and mutual coupling are always in cooperation for any $N$ in this case. The cooperation can also be seen numerically from Fig.~\hyperref[fig3]{3{\bf b}} where we observe scaled synchronization measure $\mathbb{S}_{max} \equiv (2\pi)^N S_{max}$ always exceeds the contribution from entrainment for any $N$ even if we relax the assumption of homogeneous driving. Moreover, as $N\rightarrow\infty$, the gap between $\mathbb{S}_{max}$ and entrainment contribution grows which means mutual coupling is the dominant synchronization mechanism in the large $N$ limit. This is evident since the number of terms contributing to the degenerate coherence ($\tilde{\rho}_{jk}$) scale as $N^2$ while those contributing to non-degenerate coherences ($\tilde{\rho}_{1j}$) scale linearly with $N$. Additionally, the normalization of the density matrix induces an additional $N^{-2}$ scaling for all coherences [see Eqs.~\eqref{eq:ss_nondeg_coherences}-\eqref{eq:ss_deg_coherences}]. Thus, overall we predict that in the refrigerator regime, in the limit of $N\rightarrow\infty$, $\mathbb{S}_{max}$ is mutual coupling dominant (see \hyperref[appendixC]{Append. C}) and reads,
\begin{eqnarray}
    \label{eq:smax_asymptot}
    \mathbb{S}_{max} & = & \lim_{N\rightarrow\infty}(2\pi)^N S_{max} \nonumber \\
    &\substack{=\\n_c>n_h}&\frac{\gamma_c(n_c-n_h)}{8n_h[\gamma_c(1+n_c)+\gamma_h(1+n_h)]}.
\end{eqnarray}
The asymptotic scaled $\mathbb{S}_{max}$ above only depends on the bath properties and is independent of the drive strength. Furthermore, as shown in Fig.~\hyperref[fig3]{3{\bf b}} $S_{max}$ follows a sublinear power law behavior and all the optimum phases $\{\varphi_{j1}\}|_{S = S_{max}}$ coalesce to a single phase $3\pi/2$ (Fig.~\hyperref[fig3]{3{\bf d}}).

In the engine case, it is difficult to find an analytic closed-form expression for $S_{max}$. However, we numerically observe in  Fig.~\hyperref[fig3]{3{\bf a}}, that the competition between entrainment and mutual coupling persists for any $N$ since $
\mathbb{S}_{max}$ is smaller than entrainment contribution causing $\mathbb{S}_{max} \rightarrow 0$. This decay is due to phase repulsiveness because of mutual coupling as shown in \hyperref[fig3]{3{\bf c}}. Thus, in the large $N$-limit, the qualitative behavior of this model is analogous to the Kuramoto model with phase-repulsive coupling, where the mean-field synchronization order parameter approaches zero~\cite{tsimring2005repulsive}.

\noindent \textit{Summary.--} We have shown that there exists an interplay between entrainment and mutual coupling in a collectively driven-dissipative degenerate thermal maser. The interplay depends on the thermodynamic functionality of the maser, i.e., they compete in the engine regime and cooperate in the refrigerator regime. The results rely on two key ingredients: i. a coherent drive that collectively couples to the degenerate manifold causing entrainment and mutual coupling to coexist and ii. a dissipative mechanism that causes a population inversion between the non-degenerated and degenerated manifolds to observe the competition.

We demonstrate our findings using a minimal model of a generalized Scovil--Schulz-DuBois maser heat engine and show that in the thermodynamic limit ($N\rightarrow \infty$) the dominance of mutual coupling leads to phase repulsiveness causing the engine' working substance to be asynchronized ($\mathbb{S}_{max} = 0$). On the other hand, since there is cooperation in the refrigerator case, the phases coalesce to $3\pi/2$ giving a finite $\mathbb{S}_{max}$ that is independent of system properties. In other words, as the system size increases in order for the working substance to be synchronized the external drive needs to perform work on the system. 

Our work not only contributes to the growing field of quantum synchronization by adding valuable insights when distinct synchronizing mechanisms coexist but helps understand quantum heat engines from a synchronization perspective.

\begin{acknowledgments}
\paragraph{Acknowledgments.--}
This research was supported by the Institute for Basic Science in South Korea (IBS-R024-Y2). S.V. acknowledges support from a Government of India DST-QUEST grant number DST/ICPS/QuST/Theme-4/2019. The authors would like to thank V. Singh for the useful discussions.
\end{acknowledgments}

\bibliographystyle{apsrev4-1}
\bibliography{PRL_Murtadho}
%\printbibliography

\onecolumngrid
\appendix
\section{Appendix A: Synchronization in $D$-level system}\label{appendixA}
Here, we will derive the formula for the synchronization measure $S_{max}$ for a $D$-level system from the Husimi-Q representation
\begin{equation}
    \label{eqApp:husimiQ}
    Q[\rho] = \frac{D!}{\pi^{D-1}}\braket{\alpha_D|\rho|\alpha_D},
\end{equation}
where  $\ket{\alpha_D} = \sum_{n = 1}^{D}\alpha_{n}\ket{n}$ is the $SU(D)$ coherent state \cite{nemoto2000generalized} with
\begin{equation}
    \label{eqApp:SU(D)}
    \alpha_n = \begin{cases}
    e^{i\phi_{n}}\cos\theta_{n}\prod_{k =1}^{n-1} \sin\theta_{k} & 1
   \leq n<D\\
    e^{i\phi_{D}}\prod_{k=1}^{D-1}\sin\theta_k & n = D.
    \end{cases}
\end{equation}
It has been implicitly assumed that for $n = 1$ the product term is just an identity. We are only concerned with the distribution of phases, so we can integrate out the polar angles $d\Theta = 
\prod_{l = 1}^{D-1} \cos \theta_l (\sin\theta_l)^{2D-2l-1} d\theta_l $ from $Q[\rho]$ to obtain a quasi-probability distribution over a $D-1$ torus, i.e.,
\begin{equation}
     \int_{0}^{\pi/2}Q[\rho] \; d\Theta = \frac{D!}{\pi^{D-1}}\sum_{n,m = 1}^{D-1}\rho_{nm}\int_{0}^{\pi/2} \alpha_{n}^{*}\alpha_{m}\prod_{l = 1}^{D-1}\cos\theta_l\; (\sin\theta_l)^{2D-2l-1}\; d\theta_l. 
\end{equation}
The diagonal contribution ($n=m$) gives, 
\begin{equation}
 \int_{0}^{\pi/2}|\alpha_{n}|^2\prod_{l=1}^{D-1}\cos\theta_l(\sin\theta_l)^{2D-2l-1} \; d\theta_l = \frac{1}{2^{D-1}D!} \qquad \forall
     n = 1,\cdots,D,
\end{equation}
while the off-diagonal ($n 
\neq m$) contribution yields,
\begin{equation}
    \int_{0}^{\pi/2}\alpha_n^*\alpha_m\prod_{l=1}^{D-1}\cos\theta_n(\sin\theta_l)^{2D-2l-1}\; d\theta_l =\frac{\pi}{2^{D+1}D!}e^{i(\phi_{m}-\phi_{n})} \quad \quad \forall n \neq m = 1,
    \ldots, D.
\end{equation}
Therefore, combining the diagonal and the off-diagonal contributions we obtain,
\begin{equation}
    \int_{0}^{\pi/2}Q[\rho]\; d\Theta = \frac{1}{(2\pi)^{D-1}}+\frac{1}{2^{D+1}\pi^{D-2}}\sum_{n\neq m} \rho_{nm}e^{i(\phi_m-\phi_n)}.
\end{equation}
The first term represents the contribution from a uniform distribution, which can be eliminated by defining the phase quasi-probability distribution
\begin{equation}
    \label{eq:PHASE_DIST}
    S(\phi_1,\ldots, \phi_{D-1}) := \int_{0}^{\pi/2}Q[\rho]\; d\Theta - \frac{1}{(2\pi)^{D-1}} = \frac{1}{2^{D+1}\pi^{D-2}}\sum_{n\neq m}^{D}\rho_{nm}e^{i(\phi_m-\phi_n)} .
\end{equation}
The synchronization measure $S_{max}$ in the main text is simply the maximum of Eq.~\eqref{eq:PHASE_DIST}, i.e.,
\begin{equation}
    S_{max}  \equiv \max_{\phi_1, \cdots, \phi_{D-1}} S(\phi_1, \cdots, \phi_{D-1}) \leq \frac{1}{2^{D}\pi^{D-2}}C_{l_1},
\end{equation}
where $C_{l_1} = \sum_{n<m}|\rho_{nm}|$ is the $l_1$-norm of coherence \cite{jaseem2020generalized, baumgratz2014quantifying}. The upper-bound can be obtained by simply noting that each term in the summation of Eq.~\eqref{eq:PHASE_DIST} is upper-bounded by $|\rho_{nm}|$.
%%%%%%%%%%%%%%%%%%%%%%%%%%%%%%%%%%%%%%%%%%%%%%%%%%%%%%%%
\section{Appendix B: Steady-state of a generalized Scovil--Schulz-DuBois maser}\label{appendixB}
The quantum master equation~\eqref{eq:rot_QME} can be expanded into a series of linear first-order differential equations for each density matrix element. We divide the elements into three groups: populations ($\rho_{ii}$ with $i=0,\cdots,N+1$), non-degenerate coherences ($\rho_{1k}$ with $k = 2,\cdots,N+1$), and degenerate coherences ($\rho_{lk}$ with $k,l = 2,\cdots,N+1$ and $k\neq l$). We begin with the equations for the populations,
\begin{eqnarray}
    \label{eq:rho_11}
    \frac{d\R_{11}}{dt} &=& i\lambda\sum_{j=2}^{N+1}(\R_{1j} - \R_{j1})-2\gamma_c (1+n_c)\R_{11} -2\gamma_c n_c\sum_{j=1}^{N+1}\R_{jj}+2\gamma_c n_c, \\
    \label{eq:rho_kk}
    \frac{d\R_{kk}}{dt} &=& -i\lambda(\R_{1k}-\R_{k1})-2\gamma_h(1+n_h)\R_{kk} -2\gamma_h n_h\sum_{j=1}^{N+1}\R_{jj}+2\gamma_h n_h,
\end{eqnarray}
where we have eliminated $\R_{00}$ using the trace preserving condition $\R_{00} = 1-\sum_{j=1}^{N+1}\R_{jj}$. The equations for the non-degenerate and degenerate coherences read,
\begin{eqnarray}
\label{eq:rho_1k}
\frac{d\R_{1k}}{dt} &=& -(\gamma_h(1+n_h)+\gamma_c(1+n_c))\R_{1k} + i\lambda\left(\R_{11} - \sum_{j=2}^{N+1}\R_{jk}\right),\\
    \label{eq:rho_kl}
    \frac{d\R_{kl}}{dt} &=& -2\gamma_h(1+n_h)\R_{kl}-i\lambda(\R_{1l} - \R_{k1}).
\end{eqnarray}
Since we are interested in the steady state we set $d\R_{ij}/dt = 0$ and using Eq.~\eqref{eq:rho_1k} evaluate $d\R_{1k}/dt - d\R_{1l}/dt = 0$~($ k\neq l$) to obtain the relation,
\begin{equation}
    \label{eq:1k_minus_1l}
    \left[\gamma_h(1+n_h)+\gamma_c(1+n_c)\right]\left(\Rss_{1k} - \Rss_{1l}\right) = i\lambda\sum_{j=2}^{N+1}(\Rss_{jl}-\Rss_{jk}) \quad \quad \quad \quad (k\neq l).
\end{equation}
We also compute $d\R_{jl}/dt - d\R_{jk}/dt = 0$ with $j\neq l \neq k$ using Eq.~\eqref{eq:rho_kl} to obtain
\begin{equation}
    \label{eq:jl_minus_jk}
    \Rss_{jl}- \Rss_{jk} = \frac{i\lambda}{2\gamma_h(1+n_h)}(\Rss_{1k}-\Rss_{1l}).
\end{equation}
Combining Eqs.~\eqref{eq:jl_minus_jk} and \eqref{eq:1k_minus_1l} we obtain the steady-state coherences,
\begin{eqnarray}
    \label{eq:1l_equals_1k}
    \Big[\gamma_h(1+n_h)+\gamma_c(1+n_c) + \frac{iN\lambda}{2\gamma_h(1+n_h)}\Big](\Rss_{1k} - \Rss_{1l}) &=& 0,\\ \implies \Rss_{1l} &=& \Rss_{1k} \quad \quad (k\neq l).
\end{eqnarray}
We substitute the above result in Eq.~\eqref{eq:1k_minus_1l} to obtain
\begin{equation}
    \label{eq:kl_1k}
    \Rss_{kl} = \frac{\lambda}{\gamma_h(1+n_h)}\text{Im}(\Rss_{1k}).
\end{equation}
Since $\Rss_{1k} = \Rss_{1l}$ we can infer from the above that $ \Rss_{kl} = \Rss_{lk}$ for $k\neq l$ which means all $\Rss_{kl}$ are real. Substituting Eq.~\eqref{eq:kl_1k} to Eq.~\eqref{eq:1k_minus_1l} we obtain,
\begin{eqnarray}
    \label{eq:real_1k_0}
    -2(\gamma_h(1+n_h)+\gamma_c(1+n_c))\text{Re}(\Rss_{1k}) &=& 0, \\ \implies \text{Re}(\Rss_{1k}) &=& 0. 
\end{eqnarray}
This clearly implies that all $\Rss_{1k}$ are imaginary. Next, from Eq. \eqref{eq:rho_kk} we may calculate $d\R_{kk}/dt - d\R_{ll}/dt$ ($k\neq l$) and making use of Eq. \eqref{eq:1l_equals_1k} gives,
\begin{eqnarray}
    \label{eq:kk_equals_ll}
    -2\gamma_h(1+n_h)(\Rss_{kk}-\Rss_{ll}) &=& 0, \\\
    \implies \Rss_{kk} &=& \Rss_{ll} \quad \quad \quad (k\neq l)
\end{eqnarray}
Utilizing Eqs.~\eqref{eq:1l_equals_1k}-\eqref{eq:kk_equals_ll} to simplify Eq.~\eqref{eq:rho_11} results in
\begin{equation}
    \label{eq:reduced_lin_eq1}
    i\lambda N \Rss_{1k} -\gamma_c(1+2n_c)\Rss_{11} - N\gamma_c n_c\Rss_{kk} + \gamma_c n_c = 0.
\end{equation}
Similarly from Eqs.~\eqref{eq:rho_kk} and~\eqref{eq:rho_1k} we obtain,
\begin{eqnarray}
    \label{eq:reduced_lin_eq2}
    -i\lambda \Rss_{1k} -\gamma_h(1+n_h(1+N)) \Rss_{kk}-\gamma_h n_h\Rss_{11} +\gamma_h n_h &=& 0, \\
     \label{eq:reduced_lin_eq3}
    -\Big[\gamma_h(1+n_h)+\gamma_c(1+n_c)+\frac{\lambda^2(N-1)}{\gamma_h(1+n_h)}\Big]\Rss_{1k} + i\lambda\Rss_{11} -i\lambda\Rss_{kk} &=& 0.
\end{eqnarray}
Solving Eqs. \eqref{eq:reduced_lin_eq1}-\eqref{eq:reduced_lin_eq3} simultaneously gives the final solution
\begin{equation}
    \label{eq:nondeg_coh}
    \Rss_{1k} = \frac{i\lambda(n_c-n_h)(1+n_h)\gamma_c\gamma_h}{F(N, \lambda, \gamma_c,\gamma_h, n_h, n_c)},
\end{equation}
where $F(N, \lambda, \gamma_c,\gamma_h, n_h, n_c) = AN^2 + BN + C$, with
\begin{eqnarray}
    A &=& \lambda^2 n_h(\gamma_c(1+n_c)+\gamma_h(1+n_h)), \\
    \label{eq:B}
    B &=& \lambda^2[\gamma_c(1+3n_c+2n_h n_c) + \gamma_h (1+n_h)(1+2n_h)]+n_h\gamma_h\gamma_c(1+n_h)(1+n_c)[\gamma_h(1+n_h)+\gamma_c(1+n_c)], \\
    \label{eq:C}
    C &=& \gamma_h\gamma_c(1+n_h)^2(1+2n_c)(\gamma_h(1+n_h) +\gamma_c(1+n_c)).
\end{eqnarray}
After obtaining \eqref{eq:nondeg_coh}, it is only a matter of substitution to solve for the other steady-state density matrix elements that read
\begin{eqnarray}
    \label{eq:degcoh}
    \Rss_{jl} &=& \frac{\lambda^{2}\gamma_c(n_c - n_h)}{F(N, \lambda, \gamma_c, \gamma_h, n_h, n_c)},\\
    \label{eq:rho11}
    \Rss_{11} &=& \frac{N\lambda^2(1+n_h)(n_h\gamma_h +n_c\gamma_c)+\gamma_c\gamma_h n_c(1+n_h)^2(\gamma_c(1+n_c)+\gamma_h(1+n_h))}{F(N, \lambda, \gamma_c, \gamma_h, n_h, n_c)},\\
    \Rss_{jj} &=& \frac{(N\lambda^2 n_h+\gamma_c\gamma_h n_h(1+n_h)(1+n_c))(\gamma_c(1+n_c)+\gamma_h(1+n_h))+\lambda^2(n_c-n_h)}{F(N, \lambda, \gamma_c, \gamma_h, n_h, n_c)},\\
    \Rss_{00} &=& 1- \Rss_{11} -\sum_{j=2}^{N+1}\Rss_{jj}, \\ 
    \Rss_{01} &=& \Rss_{0j} = 0.
\end{eqnarray}
Equations \eqref{eq:nondeg_coh}-\eqref{eq:degcoh} are the same as \eqref{eq:ss_nondeg_coherences}-\eqref{eq:ss_deg_coherences} in the main text.
%%%%%%%%%%%%%%%%%%%%%%%%%%%%%%%%%%%%%%%%%%%%%%%%%%%%%%%%
\section{Appendix C: $S_{max}$ calculation for generalized Scovil--Schulz-DuBois maser}\label{appendixC}
In this section, we analytically calculate $S_{max}$ using the steady-state solution obtained in~\hyperref[appendixB]{Append. B}. We first focus on the refrigerator regime ($n_c > n_h$). In this case, we have $\arg(\rho_{1j}) = \pi/2$ and $\arg(\rho_{jl}) = 0$ ($j\neq l$) and by using the steady-state solutions [Eqs.~\eqref{eq:nondeg_coh} and \eqref{eq:degcoh}] in the phase quasi-probability distribution, Eq.~\eqref{eq:PHASE_DIST}, we obtain, 
\begin{equation}
    \label{eq:smax_refrigerator_formula}
    S_{max}|_{n_c>n_h}= \frac{1}{2^{N+2}\pi^{N}}\max_{\{\varphi_{1j}\}}\Big[\sum_{j=2}^{N+1}|\Rss_{1j}|\sin \varphi_{j1} - \sum_{j< l}^{N+1}|\Rss_{jl}|\cos(\varphi_{j1} - \varphi_{l1}) \Big] = \frac{1}{2^{N+2}\pi^N}\sum_{j
    <l}^{N+1}|\Rss_{jl}|.
\end{equation}
Above $\varphi_{ij} = \phi_i - \phi_j$. The second equality is obtained by choosing optimum phases $\varphi_{j1} = 3\pi/2\; \forall j = 2,\cdots,N+1$ that maximize $S$. Using the analytic solution obtained from Eqs.~\eqref{eq:nondeg_coh}-\eqref{eq:degcoh}, $S_{max}$ can be explicitly computed as,
\begin{equation}
    \label{eq:smax_N_refrigerator}
    S_{max}|_{n_c>n_h} = \frac{1}{(2
    \pi)^N}\frac{\lambda^2\gamma_c(n_c-n_h)(N^2+(2k-1)N)}{8F(N, \lambda, \gamma_c, \gamma_h, n_h, n_c)},
\end{equation}
where $k = \gamma_h(1+n_h)/\lambda = |\rho_{1j}^{ss}|/|\rho_{jk}^{ss}|$ is the \textit{dissipation-to-driving} ratio. In the limit of macroscopic degeneracy $N\rightarrow \infty$, $(2\pi)^N S_{max}$ approaches a constant value given in Eq.~\eqref{eq:smax_asymptot} of the main text.

In the engine case ($n_c<n_h$), the optimization is trickier. In this regime, we have competition between entrainment and mutual coupling as can be seen from $\arg(\rho_{1j}) = - \pi/2$ and $\arg(\rho_{jl}) = \pi$ [see Eqs.~\eqref{eq:nondeg_coh} and \eqref{eq:degcoh}]. The synchronization measure $S_{max}$ can be expressed as,
\begin{equation}
\label{eq:smax_engine}
        S_{max}|_{n_h>n_c} = \frac{1}{2^{N+2}\pi^N}\max_{\{\varphi_{1j}\}}\Big[\sum_{j=2}^{N+1}|\Rss_{1j}|\sin \varphi_{j1} - \sum_{j< l}^{N+1}|\Rss_{jl}|\cos(\varphi_{l1} - \varphi_{j1}) \Big].
\end{equation}
Optimization of Eq. \eqref{eq:smax_engine} is difficult for an arbitrary $N$. Let us check the simplest non-trivial case of $N=2$. In this case, $S_{max}$ can be cast into a simple form
\begin{equation}
    \label{eq:smax_engine2}
    S_{max}|_{n_h>n_c} = \frac{1}{16\pi^2} |\Rss_{23}|\max_{\varphi_{21}, \varphi_{31}}\Bigl( k\sin \varphi_{21}+ k\sin\varphi_{31}-\cos(\varphi_{31}-\varphi_{21})\Bigr).
\end{equation}
Thus, calculating $S_{max}$ is now reduced to optimizing a two-variable function $f(x,y) \equiv k\sin x + k\sin y - \cos(x-y)$. One can easily verify
\begin{equation}
    \label{eq:max_fxy}
    \max_{x,y}f(x,y) = \begin{cases}
    2k -1 & \text{if } k\geq 2\\
    1+\dfrac{k^2}{2} & \text{if } 0\leq k\leq 2,
    \end{cases}
\end{equation}
with optimum points $(x,y) = (\pi/2, \pi/2)$ when $k\geq 2$ and $\{(\arcsin(k/2), \pi-\arcsin(k/2)),(\pi-\arcsin(k/2), \arcsin(k/2))\}$ when $k\leq 2$. By substituting Eq.~\eqref{eq:max_fxy} in Eq.~\eqref{eq:smax_engine2}, one obtains Eq. \eqref{eq:smax_result} of the main text.  
%%%%%%%%%%%%%%%%%%%%%%%%%%%%%%%%%%%%%%%%%%%%%%%%%%%%%%%%
\section{Appendix D: $S_{max} = 0$ if and only if $\rho$ is diagonal ($D = 3$)}\label{appendixD}
Next, we will show that in the case of $D = 3$, $S_{max} = 0$ if and only if $\rho$ is diagonal. We consider a three-level system with $\{\ket{0}, \ket{1}, \ket{2}\}$ representing the eigenvectors. A general expression for $S\equiv S(\phi_0, \phi_1, \phi_2)$ for such a three-level system reads,
\begin{equation}
    S = \frac{1}{8\pi}[|\rho_{01}|\cos(\phi_1 - \phi_0+\Phi_{01}) + |\rho_{02}|\cos(\phi_{2}-\phi_{0}+\Phi_{02}) + |\rho_{12}|\cos(\phi_{2} - \phi_{1} +\Phi_{12})],
\end{equation}
where $\Phi_{ij} = \arg(\rho_{ij})$. We first transform the equation by defining $\varphi_{ij} = \phi_{i} - \phi_{j}$, i.e.,
\begin{equation}
    S = \frac{1}{8\pi}[|\rho_{01}|\cos(\varphi_{10}+\Phi_{01}) + |\rho_{02}|\cos(\varphi_{20}+\Phi_{02}) + |\rho_{12}|\cos(\varphi_{20} - \varphi_{10} +\Phi_{12})].
\end{equation}
Given that the reduced density matrix $\rho$ is diagonal, $S(\varphi_{10},\varphi_{20})$ is zero everywhere ($\because |\rho_{ij}|=0~\forall i,j$) and thus it is trivial that $S_{max} = 0$. 

However, it is not trivial to show that if $S_{max}=0$ the $\rho$ will be diagonal. We will prove it by contradiction. Let us assume $\rho$ is \textit{not} diagonal and $S_{max} = 0$. Then, by definition $S(\varphi_{10},\varphi_{20})$ is zero or negative \textit{everywhere else}. We will show below, considering all possible cases, that we can always find $\{\varphi_{10}, \varphi_{20}\}$ such that $S$ is positive, ergo contradiction. 
\noindent \textbf{Case 1:} Only one coherence is non-zero, let's say $\rho_{01}$. Then, we can choose $\varphi_{10} = -\Phi_{01}$ such that $S = |\rho_{01}|>0$.

\noindent \textbf{Case 2:} Two coherences are non-zero, let's say $\rho_{01}$ and $\rho_{02}$. We can then choose $\varphi_{10} = -\Phi_{01}$ and $\varphi_{20} = -\Phi_{02}$ such that $S = |\rho_{01}| + |\rho_{02}|>0$

\noindent \textbf{Case 3:} All coherences are non-zero. This is a non-trivial case. First, let us choose $(\varphi_{10}, \varphi_{20}) = (\pi/2 - \Phi_{01}, \pi/2 - \Phi_{02})$ such that
\begin{equation}
    S = \frac{1}{8\pi}|\rho_{12}|\cos(\Phi_{01}-\Phi_{02}+\Phi_{12}) > 0.
\end{equation}
The above is positive if the cosine term is positive. If it is negative, we can just choose $(\varphi_{10}, \varphi_{20}) = (-\pi/2 - \Phi_{01}, \pi/2 - \Phi_{02})$ such that $S$ remains positive, i.e.,
\begin{equation}
    S = -\frac{1}{8\pi}|\rho_{12}|\cos(\Phi_{01}-\Phi_{02}+\Phi_{12}) > 0.
\end{equation}
If the cosine term is zero, we choose $(\varphi_{10}, \varphi_{20}) = (- \Phi_{01}, - \Phi_{02})$ to keep $S$ positive,
\begin{equation}
    S = \frac{1}{8\pi}\left(|\rho_{01}| + |\rho_{02}|\right)>0.
\end{equation}
Thus, we conclude that in all cases, we can always find phases configuration such that $S$ is positive, implying that $S_{max}$ cannot be zero if $\rho$ is not diagonal. 
%%%%%%%%%%%%%%%%%%%%%%%%%%%%%%%%%%%%%%%%%%%%%%%%%%%%%%%%

\end{document}